\documentclass[aps,twocolumn,showpacs,superscriptaddress]{revtex4}
\usepackage{epsf}
\usepackage{amsmath}    

\begin{document}

\title{Search on a Fractal Lattice using a Quantum Random Walk}
\author{Apoorva Patel}
\email{adpatel@cts.iisc.ernet.in}
\affiliation{Centre for High Energy Physics,
             Indian Institute of Science, Bangalore-560012, India}
\affiliation{Supercomputer Education and Research Centre,
             Indian Institute of Science, Bangalore-560012, India}
\author{K.S. Raghunathan}
\email{ksraghu@gmail.com}
\affiliation{Centre for High Energy Physics,
             Indian Institute of Science, Bangalore-560012, India}
\date{\today}

\begin{abstract}
The spatial search problem on regular lattice structures in integer number
of dimensions $d\geq2$ has been studied extensively, using both coined and
coinless quantum walks. The relativistic Dirac operator has been a crucial
ingredient in these studies. Here we investigate the spatial search problem
on fractals of non-integer dimensions. Although the Dirac operator cannot be
defined on a fractal, we construct the quantum walk on a fractal using the
flip-flop operator that incorporates a Klein-Gordon mode. We find that the
scaling behavior of the spatial search is determined by the spectral (and
not the fractal) dimension. Our numerical results have been obtained on the
well-known Sierpinski gaskets in two and three dimensions.
\end{abstract}
\maketitle

\section{Introduction}

The spatial search problem is to find a marked object from an unsorted
database of size $N$ spread over distinct locations, with the restriction
that one can proceed from any location to only its neighbors while
inspecting the objects. Classical algorithms for the unsorted database search
can do no better than inspect one location after another, and so are $O(N)$.
On the other hand, quantum algorithms can do better by working with a
superposition of states. The quantum spatial search problem has been studied
extensively in recent years, in a variety of geometries ranging from a single
hypercube to regular lattices in various integer dimensions (see for example,
Refs. \cite{hypsrch1,gridsrch1,qwalk2,tulsi,hexsrch,dgt2search,deq2search}).
These investigations used local translationally invariant quantum walk
operators \cite{qrw,qrwrev}, arising from relativistic quantum mechanics
in some form, to obtain their results. Here we investigate the spatial
search on a fractal lattice, which has neither a translational symmetry nor
a straightforward structure to implement the relativistic Dirac operator.
The non-integral value of the fractal dimension also lets us explore the
dependence of the spatial search on the geometry of space in more generality.

Our algorithmic strategy for the quantum search is to construct a Hamiltonian
evolution, whereby the kinetic term diffuses the amplitude distribution all
over the space and the potential term attracts the amplitude distribution
toward the marked vertex \cite{grover_strategy}. In its discrete form,
the kinetic term is realized as the quantum walk operator $W$, explicitly
defined in Section III. A relativistic walk provides the fastest diffusion,
and is expected to produce the quickest search. The best potential term is
the one that provides maximum contrast between the marked vertex and the
rest. In its discrete form, it is the binary oracle. Choosing the origin
as the marked vertex,
\begin{eqnarray}
&& V = V_0 \delta_{\vec{x},0} ~,~~ e^{-iV_0 \tau} = -1 \cr
&\Longrightarrow& R = e^{-iV\tau} = I - 2 |\vec{0}\rangle\langle\vec{0}| ~,
\end{eqnarray}
where $\tau$ is the time-step size. The algorithm alternates between the
oracle and the walk operators, yielding the time evolution
\begin{equation}
|\psi(\vec{x};t_1,t_2)\rangle = [W^{t_1} R]^{t_2} |\psi(\vec{x};0,0)\rangle ~.
\label{evolsearch}
\end{equation}
Here $t_2$ is the number of oracle calls, and $t_1$ is the number of walk
steps between the oracle calls. Both have to be optimized, depending on
the spatial distribution of the vertices, to concentrate the amplitude
distribution toward the marked vertex as quickly as possible and to solve
the spatial search problem.

The iterative evolution of Eq.(\ref{evolsearch}) redistributes the amplitude
at each vertex over itself and its neighbors at every step. Unitarity of
the evolution means that the eigenvalues have unit magnitude, and so the
results of the algorithm are periodic in time. Grover's algorithm is the
special case where one can move from any vertex to any other vertex in just
one step, which can be interpreted as the $d\rightarrow\infty$ limit of
spatial search \cite{dgt2search}. Unlike Grover's algorithm, the maximum
probability of being at the marked vertex, $P$, does not reach the value
$1$ for a generic spatial search. Augmenting the algorithm by the amplitude
amplification procedure \cite{brassard}, the marked vertex can be found
with probability $\Theta(1)$, and the overall complexity of the algorithm
is then characterized by the effective number of oracle calls $t_2/\sqrt{P}$.

We have argued \cite{dgt2search} that the spatial search in $d$ dimensions
obeys two lower bounds,
\begin{equation}
t_2 ~\geq~ {\rm max}\{d N^{1/d}, \pi\sqrt{N}/4\} ~.
\label{bounds}
\end{equation}
They, respectively, follow from the distinct physical principles of special
relativity (i.e. the walk must be able to travel across the lattice) and
unitarity (i.e. Grover's algorithm is optimal in the absence of any constraint
on walk movements). The best spatial search algorithms, therefore, appear in
the framework of relativistic quantum mechanics. Moreover, an analogy with
statistical mechanics of critical phenomena in different dimensions, having
an interplay of multiple dynamical features, provides an understanding of
the scaling behavior of the spatial search. For $d>2$, the latter bound wins,
and the spatial search is $O(\sqrt{N})$ differing only in scaling prefactors.
The two bounds cross in the critical dimension $d=2$, where logarithmic
corrections to scaling behavior are expected and have indeed been observed
\cite{gridsrch1,qwalk2,tulsi,hexsrch,deq2search}. For $d<2$, the former
bound dominates and should govern the scaling behavior of the spatial search.
But the relativistic evolution operator is power-law infrared divergent
there (as $\int d^d k / k^2$ in the continuum formulation), which may
modify the scaling behavior compared to the bound.

The preceding arguments do not provide any new insight for $d=1$, where both
classical and quantum spatial search are $O(N)$. We need non-integral values
of $d$ to convincingly test the analogy of the spatial search with critical
phenomena in statistical mechanics, and to that end we have investigated
the spatial search on fractal lattices. Many candidates for fractal lattices
are available with $1<d<2$, and for them we expect the spatial search to have
a power-law scaling form, say $O(N^s)$. We want to determine how the exponent
$s$ depends on $d$, in particular whether it indeed equals $1/d$, and we
specifically work with fractals known as Sierpinski gaskets.

\section{Quantum Walk\break in a Fractal Geometry}

Fractals are self-similar structures, and are conveniently defined using a
recursive scheme in an embedding space, say of Euclidean dimensions $d_E>d$.
(This is in contrast to regular lattices generated by repetitive unit cells
in translationally invariant systems, where $d_E=d$.) For Sierpinski gaskets,
the recursive structure is a regular simplex. That is an equilateral triangle
for $d_E=2$, and Fig.~1 illustrates the recursive generation of the Sierpinski
gasket as a function of the stage number $S$.

\begin{figure}[b]
\epsfxsize=8cm
\hspace{-5mm}\epsfbox{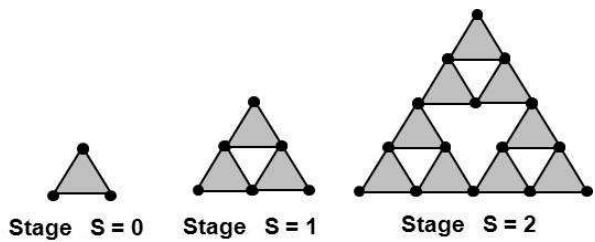}
\caption{Sierpinski gasket generation at various stages.}
\end{figure}

At stage $S$, the linear extent of Sierpinski gaskets is $L=2^S$.
Since a simplex has $d_E+1$ vertices, the total number of vertices $N$
in Sierpinski gaskets is given by,
\begin{eqnarray}
N(S) &=& \frac{(d_E+1)((d_E+1)^S+1)}{2} \cr
     &\mathop{\longrightarrow}\limits^{S\rightarrow\infty}& 
     \left(\frac{d_E+1}{2}\right) L^{\ln(d_E+1)/\ln 2} ~.
\end{eqnarray}
Thus the fractal (also called the Hausdorff) dimension of Sierpinski
gaskets is
\begin{equation}
d = \ln(d_E+1)/\ln 2 ~.
\end{equation}

Dynamical phenomena on fractals have been investigated before
\cite{AO:JPHY,RR:JPHY}, in particular the classical random walk and the
vibrational problem. Instead of the translational symmetry of regular
lattices, self-similar fractals have the dilation (or the scale) symmetry.
That leads to power-law scaling behavior for physical observables, and
the critical exponents for various well-known fractals have been calculated.
The most common operator appearing in the dynamical equations is the
Laplacian, and the critical behavior is then governed by the scaling of
its spectral density,
\begin{equation}
\rho(\omega) \sim \omega^{d_s-1} ~.
\end{equation}
This result can be understood as the reciprocal space for the Laplacian
(or the Brillouin zone) having an effective dimension $d_s$, which is
called the spectral (or the fracton) dimension \cite{AO:JPHY,RR:JPHY}.
In the case of Sierpinski gaskets,
\begin{equation}
d_s = 2\ln(d_E+1)/\ln(d_E+3) ~,
\end{equation}
which cannot exceed $2$ although $d$ can.

For regular lattices, $d_E = d = d_s$, whereas it is observed that
$d_E \geq d \geq d_s$ in general. Note that both the position and
the reciprocal space describe the same number of degrees of freedom
in the same number of embedding dimensions. So $d \ne d_s$ means that
the arrangement of vertices is different (in terms of connectivity
as well as spacings) in the position and the reciprocal spaces.
For the Sierpinski gaskets in two and three embedding dimensions,
the relevant dimensions are:
\begin{eqnarray}
    d_E = 2 &:& ~~d   = 1.5849625\ldots ~,~~ d_s = 1.3652124\ldots ~, \cr
    d_E = 3 &:& ~~d   = 2               ~,~~ d_s = 1.5474112\ldots ~.
\end{eqnarray}

Both the classical random walk and the vibrational problem involve the
same Laplacian in their spatial dependence, although they differ in their
temporal dependence (first and second derivative respectively). So their
critical exponents depend on the same $d_s$, but the dependence differs
in phenomena involving temporal correlations. For the relativistic quantum
walk problem we are addressing, the appropriate scaling behavior corresponds
to that for the vibrational problem (and not the classical random walk).

Now quantum walk diffuses at relativistic speeds on a hypercubic lattice
when the Dirac operator is used for time evolution. But it is unclear how to
extend the relevant Clifford algebra to non-integer dimensions \cite{vielbein}.
So our earlier methodology for the spatial search on hypercubic lattices
\cite{qwalk2,dgt2search,deq2search} cannot be implemented on fractals,
and we need a different implementation of a quantum relativistic walk.
It is known that the flip-flop walk diffuses relativistically \cite{gridsrch1},
and can be easily applied to graphs where all the vertices have the same
connectivity. That happens to be the case for Sierpinski gaskets, which have
$2d_E$ neighbors for every vertex. We therefore first construct the flip-flop
walk on Sierpinski gaskets, in terms of independent walk steps along every
link, and then use it to study the spatial search problem.

\section{The Flip-Flop Walk}

The flip-flop walk \cite{gridsrch1} is constructed in the joint Hilbert
space containing both the position and the link degrees of freedom,
\begin{equation}
|\psi(\vec{x},\hat{l})\rangle \equiv \sum_{x,l} a_{x,l}
     |\vec{x}\rangle \otimes |\hat{l}\rangle
     \in {\cal H}_N \otimes {\cal H}_k \equiv {\cal H}_{\rm search} ~.
\label{genstate}
\end{equation}
Every link attached to a given vertex is assigned its own direction pointing
away from the vertex, so the dimension $k$ of ${\cal H}_k$ is the number of
neighbors for the vertex. (For ease of implementation, we take this number
to be the same for each vertex.) The walk step consists of a shift operation
followed by a mixing of the link amplitudes, i.e. $W=G{\cal S}$. The shift
operation first propagates the quantum amplitude along its link direction,
and then reverses the link direction,
\begin{equation}
|\vec{x}\rangle \otimes |\hat{l}\rangle
     \mathop{\longrightarrow}\limits^{\cal S} 
     |\vec{x}+\hat{l}\rangle \otimes |-\hat{l}\rangle ~.
\end{equation}
Note that when the graph has a link $\hat{l}$ at the vertex $\vec{x}$,
it automatically has a link $-\hat{l}$ at the vertex $\vec{x}+\hat{l}$.
The mixing of links at every vertex is the reflection in the mean
operation (also referred to as the Grover coin \cite{qrwrev}),
\begin{equation}
a_{x,l} \mathop{\longrightarrow}\limits^G
        \frac{2}{k} \sum_{l'} a_{x,l'} - a_{x,l} ~.
\end{equation}

Both ${\cal S}$ and $G$ are reflection operators, ${\cal S}^2 = 1 = G^2$,
and the walk spreads on the lattice because these two operators do not commute.
${\cal S}$ is block diagonal on the links, and its eigenstates are the
combinations $|\vec{x},\hat{l}\rangle \pm |\vec{x}+\hat{l},-\hat{l}\rangle$
for every link, with eigenvalues $\pm 1$. Explicitly,
\begin{equation}
{\cal S} = \mathop{\bigoplus}\limits_{\rm links} \big[
      (|\vec{x},\hat{l}\rangle+|\vec{x}+\hat{l},-\hat{l}\rangle)
      (\langle \vec{x},\hat{l}|+\langle \vec{x}+\hat{l},-\hat{l}|) - 1 \big] ~.
\end{equation}
On the other hand, $G$ is block diagonal on the vertices, 
\begin{equation}
G = \mathop{\bigoplus}\limits_{\rm vertices} \Big[ \frac{2}{k}
         \big(\sum_l |\vec{x},\hat{l}\rangle\big)
         \big(\sum_{l'}\langle \vec{x},\hat{l}'|\big) - 1 \Big] ~.
\end{equation}
For every vertex, it has one eigenvalue $+1$ corresponding to the isotropic
distribution, and $k-1$ eigenvalues $-1$ corresponding to other orthogonal
distributions.

The unbiased uniform superposition state,
\begin{equation}
|s\rangle = (Nk)^{-1/2} \sum_{x,l} |\vec{x},\vec{l}\rangle ~,
\end{equation}
is invariant under both ${\cal S}$ and $G$ (i.e. both operators have
eigenvalue $+1$). Furthermore, the oracle $R$ trivially commutes with $G$.

When the lattice has translational invariance, $\cal S$ can be simplified
by Fourier transforming to the momentum space, and then the eigenspectrum
of the walk can be obtained by diagonalizing a $k \times k$ matrix
describing the operator $W$ for each value of the momentum \cite{gridsrch1}.
Such an analysis, however, is not possible for a fractal.

\begin{figure}[b]
\vspace{-4mm}
\epsfxsize=8.8cm
\epsfbox{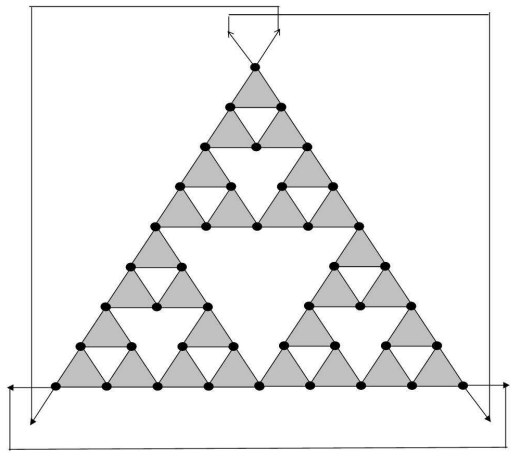}
\vspace{-8mm}
\caption{Wraparound for the corner vertices of the two-dimensional
Sierpinski gasket.}
\end{figure}

\subsection{Flip-flop walk on Sierpinski gaskets}

We first implement the flip-flop walk on the Sierpinski gasket in two
dimensions. As can be seen from Fig.~1, all the internal vertices of the
gasket have four neighbors, whereas the three corner vertices have only
two neighbors. We increase the number of neighbors for the corner
vertices to four, by providing a periodic wraparound as illustrated in
Fig.~2. Furthermore, to conveniently label the vertices and to keep track
of their neighbors, we embed the gasket in a two-dimensional rectangular
grid as shown in Fig.~3.
 
\begin{figure}
\vspace{-4mm}
\epsfxsize=8cm
\epsfbox{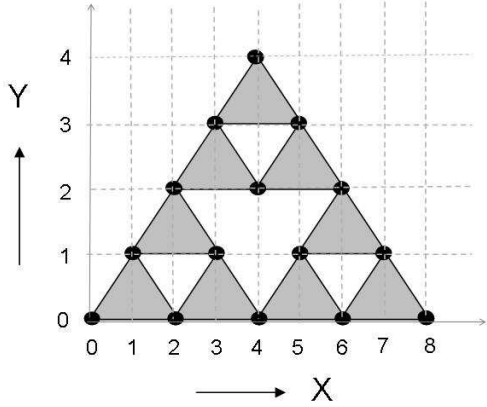}
\vspace{-4mm}
\caption{Embedding of the two-dimensional Sierpinski gasket
in a rectangular grid.}
\end{figure}

In this configuration, there are six possible walk directions in general,
as depicted in Fig.~4, out of which only four are valid for any given
vertex. We separate the vertices in to six types, also shown in Fig.~4,
depending on which of the four walk directions are applicable to them.
Note that V0, V1 and V2 are internal vertices, whereas V3, V4 and V5 are
the corner vertices.

With this infrastructure, we can now explicitly write down the operations
involved in the flip-flop walk. The shift operation is:
\begin{equation}
{\cal S} = \begin{cases}
    |x,y\rangle\otimes|0\rangle \longleftrightarrow |x+2,y  \rangle\otimes|3\rangle ~, \cr
    |x,y\rangle\otimes|1\rangle \longleftrightarrow |x+1,y+1\rangle\otimes|4\rangle ~, \cr
    |x,y\rangle\otimes|2\rangle \longleftrightarrow |x-1,y+1\rangle\otimes|5\rangle ~.
    \end{cases}
\end{equation}
The mixing of link amplitudes occurs only among the four valid directions
at each vertex. It can be described by the operator matrix,
\begin{equation}
G^{(k=4)} = 1 \otimes \frac{1}{2}\begin{pmatrix}
                    -1 & \phantom{+}1 & \phantom{+}1 & \phantom{+}1 \cr
          \phantom{+}1 &           -1 & \phantom{+}1 & \phantom{+}1 \cr
          \phantom{+}1 & \phantom{+}1 &           -1 & \phantom{+}1 \cr
          \phantom{+}1 & \phantom{+}1 & \phantom{+}1 &           -1 
                                 \end{pmatrix} ~.
\end{equation}

These definitions can be easily extended for flip-flop walks on Sierpinski
gaskets in any number of dimensions, with $k=2d_E$.

\begin{figure}
\setlength{\unitlength}{1pt}
\begin{picture}(220,250)(5,-15)
\thicklines
\put(120,180){\circle*{5}}
\put(128,182){\makebox(0,0)[bl]{$x,y$}}
\put(120,180){\vector(1,0){48}}
\put(150,172){\makebox(0,0)[bl]{\bf 0}}
\put(168,180){\circle*{5}}
\put(160,182){\makebox(0,0)[bl]{$x+2,y$}}
\put(120,180){\vector(-1,0){48}}
\put(86,172){\makebox(0,0)[bl]{\bf 3}}
\put(72,180){\circle*{5}}
\put(58,182){\makebox(0,0)[bl]{$x-2,y$}}
\put(120,180){\vector(2,3){24}}
\put(130,210){\makebox(0,0)[bl]{\bf 1}}
\put(144,216){\circle*{5}}
\put(120,180){\vector(-2,3){24}}
\put(130,222){\makebox(0,0)[bl]{$x+1,y+1$}}
\put(104,210){\makebox(0,0)[bl]{\bf 2}}
\put(66,222){\makebox(0,0)[bl]{$x-1,y+1$}}
\put(96,216){\circle*{5}}
\put(120,180){\vector(2,-3){24}}
\put(130,145){\makebox(0,0)[bl]{\bf 5}}
\put(130,130){\makebox(0,0)[bl]{$x+1,y-1$}}
\put(144,144){\circle*{5}}
\put(120,180){\vector(-2,-3){24}}
\put(104,145){\makebox(0,0)[bl]{\bf 4}}
\put(96,144){\circle*{5}}
\put(66,130){\makebox(0,0)[bl]{$x-1,y-1$}}

\put(35,60){\makebox(0,0)[bl]{V0}}
\put(40,90){\circle*{5}}
\put(40,90){\vector(1,0){20}}
\put(63,87){\makebox(0,0)[bl]{\bf 0}}
\put(40,90){\vector(2,3){10}}
\put(52,105){\makebox(0,0)[bl]{\bf 1}}
\put(40,90){\vector(2,-3){10}}
\put(52,70){\makebox(0,0)[bl]{\bf 5}}
\put(40,90){\vector(-2,-3){10}}
\put(22,70){\makebox(0,0)[bl]{\bf 4}}
\put(115,60){\makebox(0,0)[bl]{V1}}
\put(120,90){\circle*{5}}
\put(120,90){\vector(-1,0){20}}
\put(93,87){\makebox(0,0)[bl]{\bf 3}}
\put(120,90){\vector(-2,3){10}}
\put(102,105){\makebox(0,0)[bl]{\bf 2}}
\put(120,90){\vector(-2,-3){10}}
\put(102,70){\makebox(0,0)[bl]{\bf 4}}
\put(120,90){\vector(2,-3){10}}
\put(132,70){\makebox(0,0)[bl]{\bf 5}}
\put(180,60){\makebox(0,0)[bl]{V2}}
\put(185,90){\circle*{5}}
\put(185,90){\vector(1,0){20}}
\put(208,87){\makebox(0,0)[bl]{\bf 0}}
\put(185,90){\vector(-1,0){20}}
\put(158,87){\makebox(0,0)[bl]{\bf 3}}
\put(185,90){\vector(2,3){10}}
\put(197,105){\makebox(0,0)[bl]{\bf 1}}
\put(185,90){\vector(-2,3){10}}
\put(167,105){\makebox(0,0)[bl]{\bf 2}}

\put(35,-5){\makebox(0,0)[bl]{V3}}
\put(40,25){\circle*{5}}
\put(40,25){\vector(1,0){20}}
\put(63,22){\makebox(0,0)[bl]{\bf 0}}
\put(40,25){\vector(-1,0){20}}
\put(13,22){\makebox(0,0)[bl]{\bf 3}}
\put(40,25){\vector(2,3){10}}
\put(52,40){\makebox(0,0)[bl]{\bf 1}}
\put(40,25){\vector(-2,-3){10}}
\put(22,5){\makebox(0,0)[bl]{\bf 4}}
\put(115,-5){\makebox(0,0)[bl]{V4}}
\put(120,25){\circle*{5}}
\put(120,25){\vector(1,0){20}}
\put(143,22){\makebox(0,0)[bl]{\bf 0}}
\put(120,25){\vector(-1,0){20}}
\put(93,22){\makebox(0,0)[bl]{\bf 3}}
\put(120,25){\vector(-2,3){10}}
\put(102,40){\makebox(0,0)[bl]{\bf 2}}
\put(120,25){\vector(2,-3){10}}
\put(132,5){\makebox(0,0)[bl]{\bf 5}}
\put(180,-5){\makebox(0,0)[bl]{V5}}
\put(185,25){\circle*{5}}
\put(185,25){\vector(2,3){10}}
\put(197,40){\makebox(0,0)[bl]{\bf 1}}
\put(185,25){\vector(-2,3){10}}
\put(167,40){\makebox(0,0)[bl]{\bf 2}}
\put(185,25){\vector(-2,-3){10}}
\put(167,5){\makebox(0,0)[bl]{\bf 4}}
\put(185,25){\vector(2,-3){10}}
\put(197,5){\makebox(0,0)[bl]{\bf 5}}
\end{picture}
\caption{Above: possible walk directions, and below: various vertex types,
for the two-dimensional Sierpinski gasket.}
\end{figure}
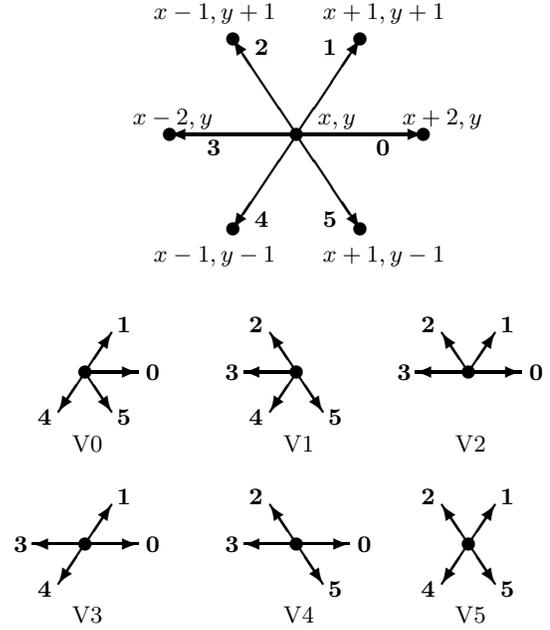

\subsection{Flip-flop walk as a relativistic propagator}

To understand the time evolution provided by the flip-flop walk operator,
we describe it for a hypercubic lattice in an integer number of dimensions,
in a notation that identifies ${\cal H}_k$ with internal degrees of freedom.

For $d=1$, the shift operator is
\begin{eqnarray}
{\cal S} &=& \sum_x (1\otimes\sigma_1) \cr
         &\times& \left[ |x+1\rangle\langle x| \otimes \frac{1+\sigma_3}{2}
                       + |x-1\rangle\langle x| \otimes \frac{1-\sigma_3}{2}
                 \right] \cr
         &=& \sum_x \Big[ |x+1\rangle\langle x| \otimes \sigma_-
                        + |x-1\rangle\langle x| \otimes \sigma_+ \Big] ~,
\end{eqnarray}
and the link-mixing operator is $G^{(k=2)}=1\otimes\sigma_1$. Together
they produce a directed walk, where the two spinor components propagate
independently and at full speed in opposite directions,
\begin{equation}
W = \sum_x \left[ |x+1\rangle\langle x| \otimes \frac{1+\sigma_3}{2}
                + |x-1\rangle\langle x| \otimes \frac{1-\sigma_3}{2} \right] ~.
\end{equation}
This is a representation of the massless Dirac propagator, in the basis
where the two spinor components describe the two chiral degrees of freedom
in $d=1$. It is also the representation of the massless scalar Klein-Gordon
propagator, where the two parts of the d'Alembert solution travel
independently in opposite directions.

The situation in higher dimensions is more non-trivial. For a hypercubic
lattice in $d$ dimensions, size $L=N^{1/d}$ and connectivity $k=2d$.
The matrix $W=G{\cal S}$ is unitary, with eigenvalues of unit magnitude
and orthogonal eigenvectors. Its spectrum can be obtained as follows.
The momentum components on a periodic lattice are $k_i = 2\pi l_i/L$,
with $l_i \in \{0,1,...,L-1\}$. Select a specific set of momentum
components $\{k_i\}$. Then pick any two-dimensional subspace, from the
$(d-1)$ linearly independent choices $(x_1,x_2),(x_1,x_3),\ldots,(x_1,x_d)$.
In this subspace,
\begin{equation}
{\cal S} = \begin{pmatrix}
           0        &  e^{-ik_a}  &  0        &  0         \cr
           e^{ik_a} &  0          &  0        &  0         \cr
           0        &  0          &  0        &  e^{-ik_b} \cr
           0        &  0          &  e^{ik_b} &  0
           \end{pmatrix} ~,
\end{equation}
and $G_{ij} = -\delta_{ij} + 1/d$. Corresponding to each direction,
${\cal S}$ has eigenvalues $\lambda = \pm 1$ with eigenvectors
$(1,\lambda e^{ik_a})$. From the degenerate pair of eigenvectors in
every two-dimensional subspace, one can construct combinations that
are orthogonal to $(1,1,1,1)$. Acting on these combinations, $G$
reduces to $-I$, and so the combinations are eigenvectors of $W$
with eigenvalues $-\lambda$. Explicitly, they are
$( 1+\lambda e^{ik_b},   \lambda e^{ik_a}(1+\lambda e^{ik_b}),
 -(1+\lambda e^{ik_a}), -\lambda e^{ik_b}(1+\lambda e^{ik_a}) )$,
with vanishing components outside the two-dimensional subspace.
The $(d-1)$ linearly independent choices of two-dimensional subspaces
thus yield $(d-1)$-fold degenerate eigenvalues $\pm 1$ for $W$. 
The remaining two eigenvalues of $W$ are a complex conjugate pair
$e^{\pm i\omega(k_i)}$, and ${\rm Tr}(W)$ determines their real part
as $\cos(\omega(k_i)) = \frac{1}{d}\sum_i\cos(k_i)$.

We observe that the $\pm 1$ eigenvalues of the flip-flop walk correspond
to local modes that can be restricted to two-dimensional subspaces,
whereas the complex conjugate pair of eigenvalues provides relativistic
propagation with speed $1/\sqrt{d}$. With this break up of the evolution
modes, we can surmise that the flip-flop walk contains a single first-order
form \cite{firstorder} of the complex massless Klein-Gordon operator, which
makes its dynamics relativistic. The scalar Klein-Gordon operator can be
mathematically constructed in any number of dimensions, even non-integer
ones, unlike the spinor Dirac operator. Our simulation results in the next
section demonstrate that the flip-flop walk indeed achieves the desired
goals for the spatial search problem, and it would be interesting to
explore its usefulness in other relativistic problems involving scalar
fields in non-integer dimensions (or curved spaces).

\section{Spatial Search\break using the Flip-Flop Walk}

We first simulated the spatial search algorithm, for a single marked
vertex on Sierpinski gaskets, as defined by Eq.(\ref{evolsearch}).
We chose the initial state to be the unbiased uniform superposition state,
as is customary. Our numerical results are described later in the section.
In particular, we needed $t_1=2$ for the search to succeed with reasonable
probability ($t_1=1$ did not work as well, and $t_1=3$ was far worse), and
both the success probability and the number of oracle calls showed power-law
dependence on the database size $N$. Note that our spatial search algorithm
is different from that of Ref.\cite{gridsrch1} for two reasons (although we
use the same flip-flop walk): (i) we allow for values of $t_1\ne1$, and
(ii) our oracle flips only the sign of the marked vertex amplitude, whereas
Ref.\cite{gridsrch1} applies the link-mixing operator $G^{(k)}$ at the
marked vertex together with the sign flip.

Since the flip-flop walk corresponds to a massless evolution propagator,
this algorithm suffers from infrared divergence and is not optimal.
Tulsi showed how to eliminate the infrared divergence of the spatial search
algorithm by controlling the evolution operators using an ancilla qubit
\cite{tulsi}. The ancilla control traps the quantum walk at the marked
vertex and enhances the algorithm's success probability. Explicitly,
in the joint space ${\cal H}_{\rm ancilla}\otimes{\cal H}_{\rm search}$,
the $|1\rangle\otimes |\vec{x},\hat{l}\rangle$ states are the original search
space states, the $|0\rangle\otimes |\vec{x}=t,\hat{l}\rangle$ state is the
new trap state that develops a nonzero amplitude, and the $|0\rangle\otimes
|\vec{x}\ne t,\hat{l}\rangle$ states maintain zero amplitudes. This feature
can be interpreted as the introduction of an effective mass at the marked
vertex in the Hamiltonian \cite{deq2search}, or as the introduction of a
self-loop at the marked vertex in the graph \cite{KMOR}. Tulsi's scheme is
illustrated in Fig.~5, where the ancilla operators are
\begin{equation}
X_\delta = \begin{pmatrix} \cos\delta & \sin\delta \cr
                          -\sin\delta & \cos\delta \end{pmatrix} ~,~~
\overline{Z} = \begin{pmatrix} -1 & 0 \cr
                                0 & 1 \end{pmatrix} ~,
\end{equation}
and the algorithm evolves the initial state $|1\rangle\otimes|s\rangle$
to the target state $|\delta\rangle\otimes|t,\hat{l}\rangle$ with
$|\delta\rangle = X_\delta^\dag|1\rangle$.
Note that for the generic state corresponding to Eq.(\ref{genstate}),
the probability of being at the marked vertex is
$\sum_l ( |a_{\vec{x}=t,l}^{(0)}|^2 + |a_{\vec{x}=t,l}^{(1)}|^2 )$,
where the superscript refers to the ancilla value.

\begin{figure}
\setlength{\unitlength}{1pt}
\begin{picture}(250,100)(10,0)
\thicklines
\put(33,25){\makebox(0,0)[r]{$|s\rangle$}}
\put(33,65){\makebox(0,0)[r]{$|1\rangle$}}
\put(234,25){\makebox(0,0)[l]{$|t\rangle$}}
\put(234,65){\makebox(0,0)[l]{$|\delta\rangle$}}
\put(35,65){\line(1,0){30}}
\put(90,65){\line(1,0){30}}
\put(145,65){\line(1,0){30}}
\put(200,65){\line(1,0){32}}
\put(65,53){\framebox(25,24){$X_{\delta}$}}
\put(105,65){\circle*{6}}
\put(120,53){\framebox(25,24){$X_{\delta}^{\dagger}$}}
\put(160,65){\circle*{6}}
\put(175,53){\framebox(25,24){$\overline{Z}$}}
\put(35,25){\line(1,0){55}}
\put(120,25){\line(1,0){25}}
\put(175,25){\line(1,0){57}}
\put(90,13){\framebox(30,24){$R$}}
\put(105,62){\line(0,-1){25}}
\put(145,13){\framebox(30,24){$W \cdot W$}}
\put(160,62){\line(0,-1){25}}
\thinlines
\put(50,8){\dashbox{5}(167,76){}}
\put(134,90){\makebox(0,0)[b]{{Iterate $Q_{\delta}$ times}}}
\end{picture}
\caption{Logic circuit diagram for Tulsi's controlled quantum spatial
search algorithm. $R$ and $W$ are the binary oracle and the quantum
walk operator respectively. We use the generalization with two walk
steps after each oracle.}
\end{figure}
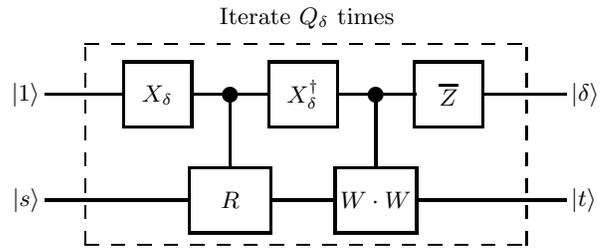

For $\delta=0$, Tulsi's algorithm reduces to spatial search described by
Eq.(\ref{evolsearch}). It finds the marked vertex with probability
$P_0=\Theta(N^{-a})$ using $Q_0=\Theta(N^b)$ oracle calls. Tulsi showed
that, with $\cos\delta = \Theta(\sqrt{P_0})$, the algorithm increases the
probability of finding the marked vertex to $P_\delta=\Theta(1)$ without
changing the scaling of oracle calls $Q_\delta=\Theta(N^b)$ \cite{tulsi}.
More explicitly, the algorithm largely confines the evolution of the
quantum state to a two-dimensional subspace of ${\cal H}_N$, whereby
\begin{equation}
P_\delta = \frac{1}{B_\delta^2} ~,~~
Q_\delta = \frac{\pi B_\delta \sqrt{N}}{4\cos\delta} ~,
\label{PQvalues}
\end{equation}
\begin{equation}
B_\delta^2 = 1+(B^2-1)\cos^2\delta ~.
\label{Bdelta}
\end{equation}
Here $B \equiv B_0 = \Theta(N^{a/2})$ is a second moment constructed from
the eigenspectrum of $W$, and characterizes the infrared divergence of the
problem. Also, Eq.(\ref{PQvalues}) relates the scaling exponents,
\begin{equation}
a = 2b-1 ~.
\label{relexp}
\end{equation}

The optimal value of the ancilla control parameter is obtained by
minimizing the algorithmic complexity,
\begin{equation}
(\cos\delta)_{\rm opt} = (B^2-1)^{-1/2} \approx 1/B ~,~~
(B_\delta^2)_{\rm opt} = 2 ~,
\label{optdelta}
\end{equation}
\begin{equation}
\Big(\frac{Q_\delta}{\sqrt{P_\delta}}\Big)_{\rm min}
= \frac{\pi\sqrt{N} B}{2} = \Theta(N^b) ~.
\label{optcomplexity}
\end{equation}
In what follows, we check these expectations regarding the scaling behavior
of the spatial search, for Sierpinski gaskets in two and three dimensions.

\subsection{Simulation results for $d_E=2$}

We simulated the spatial search algorithm on the two-dimensional
Sierpinski gasket, for stage $S=4-14$ corresponding to $N=123-7174455$.
To keep the memory requirements under control, we kept track of the
irregular pattern of vertices and their neighbors using hash tables. 
We first checked that the algorithm did produce an approximately
periodic evolution for the probability at the marked vertex, and the
probability peak was reasonably sharp for the parameters we chose.
For $S=6~(N=1095)$ and the marked vertex in the center of the gasket,
a snap shot of the probability distribution is presented in Fig.~6,
and the probability evolution at the marked vertex is shown in Fig.~7.
We find that the probability evolution is smoother with ancilla control
than without it, implying that ancilla control improves the confinement
of the evolution of the quantum state to a two-dimensional subspace of
${\cal H}_N$.

\begin{figure}
\epsfxsize=9.5cm
\epsfbox{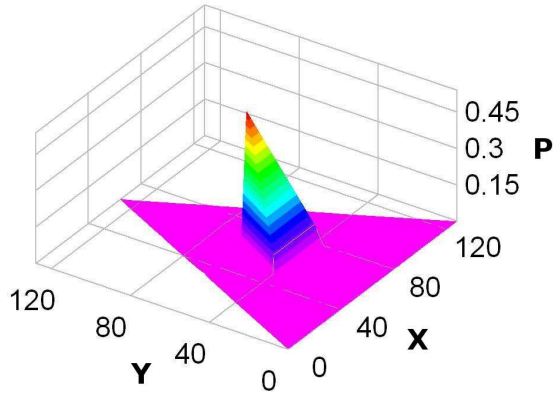}
\caption{(Color online) A snapshot of the probability distribution for
the spatial search on the two-dimensional Sierpinski gasket, when the
marked vertex attains its peak probability. The data are for $S=6$ with
ancilla control.}
\end{figure}

Implementing the spatial search algorithm without ancilla control was quite
straightforward. But to implement Tulsi's algorithm, we needed the optimal
value of $\cos\delta$, and we did not have any analytical estimate for it.
So we just assumed Eqs.(\ref{PQvalues}) and (\ref{optdelta}) to be valid,
and extracted $(\cos\delta)_{\rm opt}$ from the result for $P_0$ at each
stage $S$. That worked well, and we obtained $P_\delta \approx 0.5$ as
predicted.

Lack of translational invariance for a fractal means that the spatial
search results depend on the location of the marked vertex. In the absence
of ancilla control, each of $Q_0$, $P_0$ and $Q_0/\sqrt{P_0}$ showed
variation up to a factor of $2.2$, as we moved around the position of the
marked vertex in the gasket. With ancilla control, variation in $Q_\delta$
was somewhat smaller, up to a factor of $1.6$, whereas $P_\delta\approx0.5$
remained valid (within $10\%$) everywhere as expected. We also observed that
the largest variation in the spatial search results was between the marked
vertex positions at the corner and the center of the gasket. More importantly,
when we extracted the scaling behavior from our results, holding the
relative position of the marked vertex in the gasket (i.e. $\vec{x}/L$)
fixed, we found that the variation was essentially in the prefactors and
the scaling exponents changed by less than a percent. Henceforth we opted
to keep the marked vertex at the center of the gasket, and all our results
that follow correspond to that choice.

\begin{figure}
\epsfxsize=8.5cm
\epsfbox{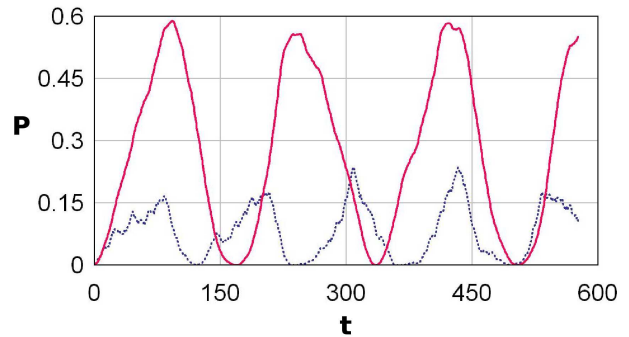}
\caption{(Color online) Time evolution of the marked vertex probability
for the spatial search on the two-dimensional Sierpinski gasket. The data
are for $S=6$. Those without ancilla control are shown as the dotted curve,
and those with ancilla control are shown as the continuous curve.}
\end{figure}

Our data for the number of oracle calls $Q$ and the corresponding peak
probability $P$ are presented in Figs.~8 and 9, both without and with
ancilla control. The figures also show simple scaling fits that were
performed in the range $S=8-14$.

In the absence of ancilla control, our fits give:
\begin{eqnarray}
\log_2 Q_0 &=&           -1.066 + 0.730\log_2 N ~ (err=0.004), \cr
\log_2 P_0 &=& \phantom{+}1.995 - 0.440\log_2 N ~ (err=0.030).
\label{PQfits2d}
\end{eqnarray}
Here ``$err$'' refers to the r.m.s. deviation in the data from the fit.
The dominant error in the asymptotic scaling parameters is the systematic 
error due to finite values of $N$. So we quote as our error estimates,
the difference in the numbers for fits corresponding to $S=7-14$ and
$S=8-14$. Then the results of Eq.(\ref{PQfits2d}) translate to the
scaling behavior:
\begin{eqnarray}
Q_0 &=& 0.478(13) ~N^{0.730(2)} ~, \cr
P_0 &=& 3.99(20)  ~N^{-0.440(4)} ~.
\end{eqnarray}
The scaling exponents satisfy the relation Eq.(\ref{relexp}) well,
which supports Tulsi's analysis criterion that the quantum state
evolution is largely confined to a two-dimensional subspace of ${\cal H}_N$.

\begin{figure}
\epsfxsize=8.8cm
\epsfbox{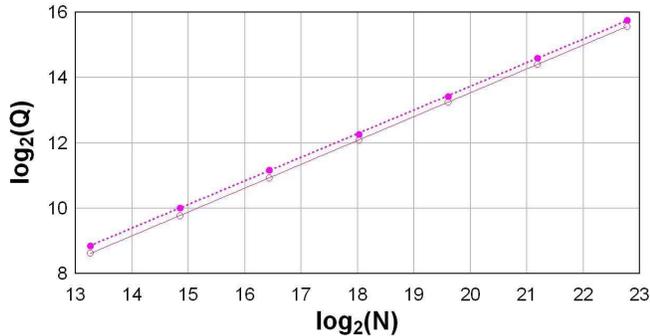}
\caption{(Color online) Scaling of the number of oracle calls for the spatial
search on the two-dimensional Sierpinski gasket. The open and filled symbols
denote data without and with ancilla control respectively. The linear fits
are for the data from $S=8$ to $S=14$.}
\end{figure}

\begin{figure}
\epsfxsize=8.8cm
\epsfbox{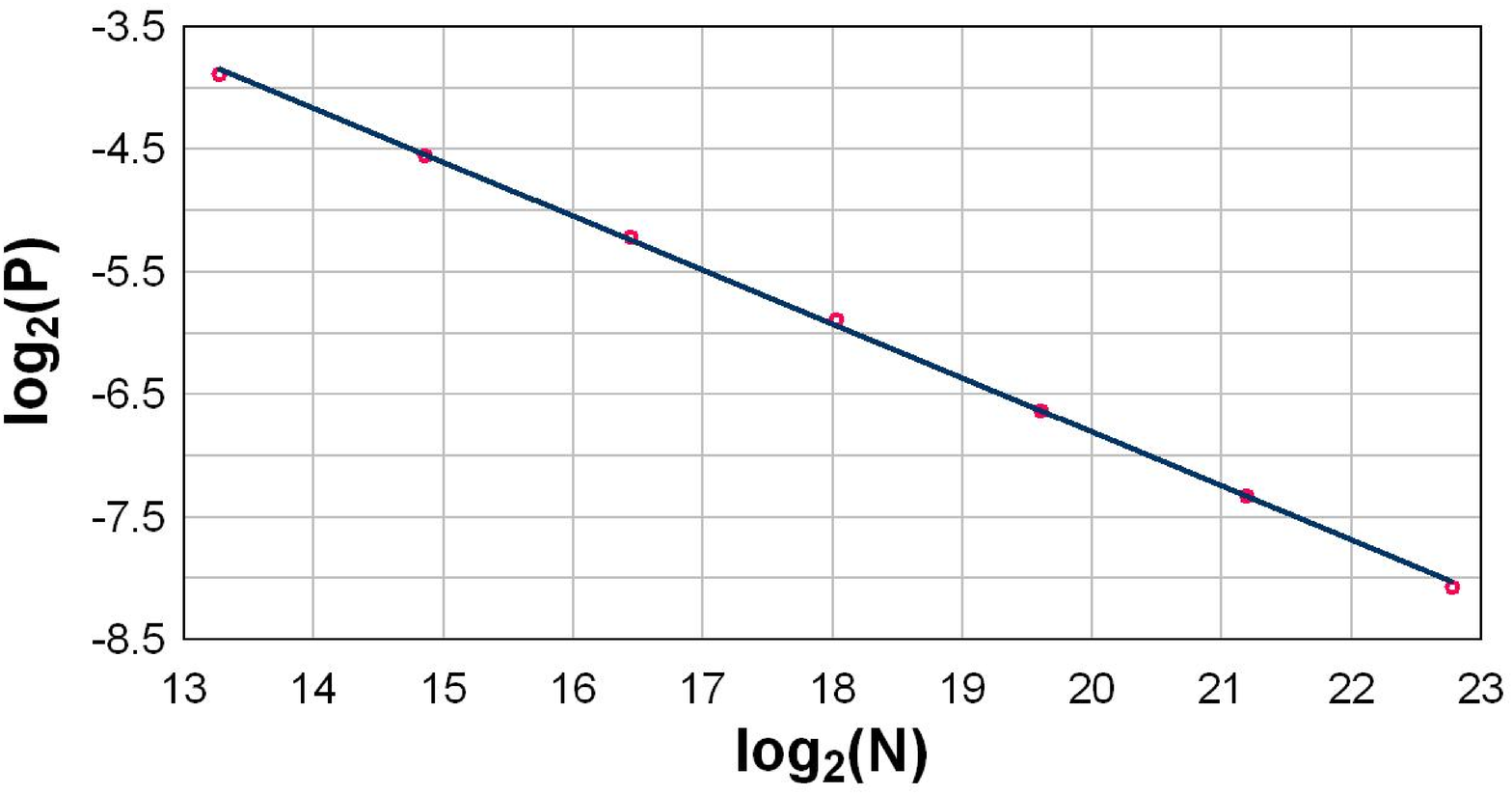}
\epsfxsize=8.8cm
\epsfbox{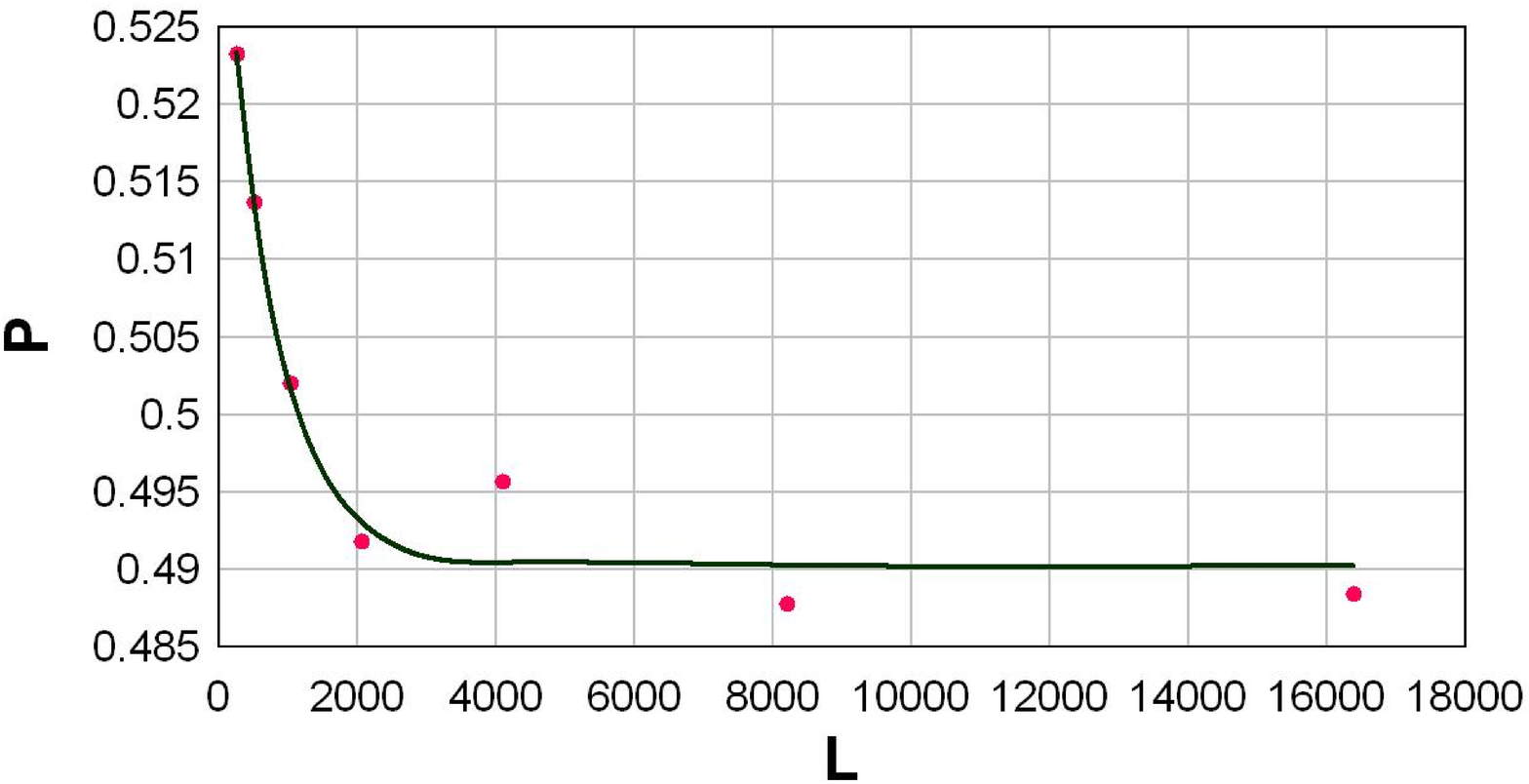}
\caption{(Color online) Scaling of the peak probability for the spatial search
on the two-dimensional Sierpinski gasket, without (top) and with (bottom)
ancilla control. The fits are for the data from $S=8$ to $S=14$, linear (top)
and exponential (bottom) respectively.}
\end{figure}

The fits for our results with ancilla control, with $\cos\delta$ determined
from the value of $P_0$, give:
\begin{eqnarray}
\log_2 Q_\delta &=& -0.739 + 0.724\log_2 N ~ (err=0.017), \\
       P_\delta &=& 0.4903 + 0.0471~e^{-0.00137L} ~ (err=0.0023). \nonumber
\end{eqnarray}
As displayed in Fig.~8, $Q_\delta$ is only marginally higher than $Q_0$,
and the scaling exponent $b$ is essentially unchanged. On the other hand,
Fig.~9 shows that $P_\delta$ approaches a constant close to $0.5$. Both
these features fully agree with Tulsi's predictions. We estimate the
scaling behavior as:
\begin{equation}
Q_\delta = 0.599(8) ~N^{0.724(1)} ~,~~ P_\delta = 0.4903(2) ~.
\end{equation}
Also, direct fits for the complexity behavior give:
\begin{eqnarray}
\log_2 (Q_\delta/\sqrt{P_\delta}) &=& -0.334 + 0.729\log_2 N ~ (err=0.015), \cr
Q_\delta/\sqrt{P_\delta} &=& 0.794(13) ~N^{0.729(1)} ~.
\end{eqnarray}

We observe that the exponent $b \approx \frac{1}{d_s} = 0.73248676\ldots$.
That clearly obeys the bound in Eq.(\ref{bounds}), since $1/d \leq 1/d_s$.
But more than that, our results imply that, for a fractal, the relevant
length scale for the spatial search is not its linear extent $L \sim N^{1/d}$,
rather it is the inverse of the linear extent of its reciprocal space
$N^{1/d_s}$. Thus it is the smaller spectral dimension which governs the
dynamics of the search process on fractals.

\subsection{Simulation results for $d_E=3$}

To reinforce our observations, we extended our simulations of the spatial
search algorithm to the three-dimensional Sierpinski gasket. In this case,
the recursive structure is a tetrahedron, whereas the linear extent at
stage $S$ remains $L=2^S$. Each internal vertex of the gasket now has six
neighbors, and every corner vertex has a periodic wraparound with the other
three corner vertices. Furthermore, there are twelve possible walk
directions in general, out of which only six are valid at any given vertex.
In our explicit operations, we used six types of internal vertices, four
types of corner vertices, and the link-mixing operator $G^{(k=6)}$.

We obtained results for stage $S=3-10$ corresponding to $N=130-2097154$,
and the marked vertex at the center of the gasket. Although not as extensive
as the results for the $d_E=2$ case, these are sufficient to discern the
patterns in the scaling laws.

We again first simulated the algorithm without ancilla control, and then
used $(\cos\delta)_{\rm opt}$ extracted from the value of $P_0$ for each
stage $S$ to simulate the algorithm with ancilla control. Our results
are plotted in Figs.~10 and 11, together with the simple scaling fits
that were performed in the range $S=5-10$. Without ancilla control,
our fits give:
\begin{eqnarray}
\log_2 Q_0 &=& -0.767 + 0.641\log_2 N ~ (err=0.002), \cr
\log_2 P_0 &=& \phantom{+}1.39 - 0.265\log_2 N ~ (err=0.036).
\end{eqnarray}
Again, ``$err$'' refers to the r.m.s. deviation in the data from the fit.
For the error estimates on the fit parameters, we quote the difference in
the numbers for fits corresponding to $S=5-10$ and $S=6-10$. Then the
corresponding estimates of the scaling behavior are:
\begin{eqnarray}
Q_0 &=& 0.588(1) ~N^{0.641(1)} ~, \cr
P_0 &=& 2.62(25) ~N^{-0.265(7)} ~.
\end{eqnarray}
Once more, the scaling exponents reasonably satisfy the relation
Eq.(\ref{relexp}), and $b \approx \frac{1}{d_s} = 0.64624063\ldots$.

\begin{figure}
\epsfxsize=8.8cm
\epsfbox{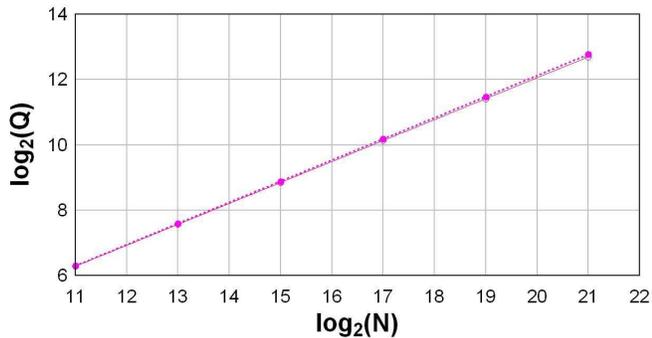}
\caption{(Color online) Scaling of the number of oracle calls for the spatial
search on the three-dimensional Sierpinski gasket. The open and filled symbols
denote data without and with ancilla control respectively. The linear fits
are for the data from $S=5$ to $S=10$.}
\end{figure}

\begin{figure}
\epsfxsize=8.8cm
\epsfbox{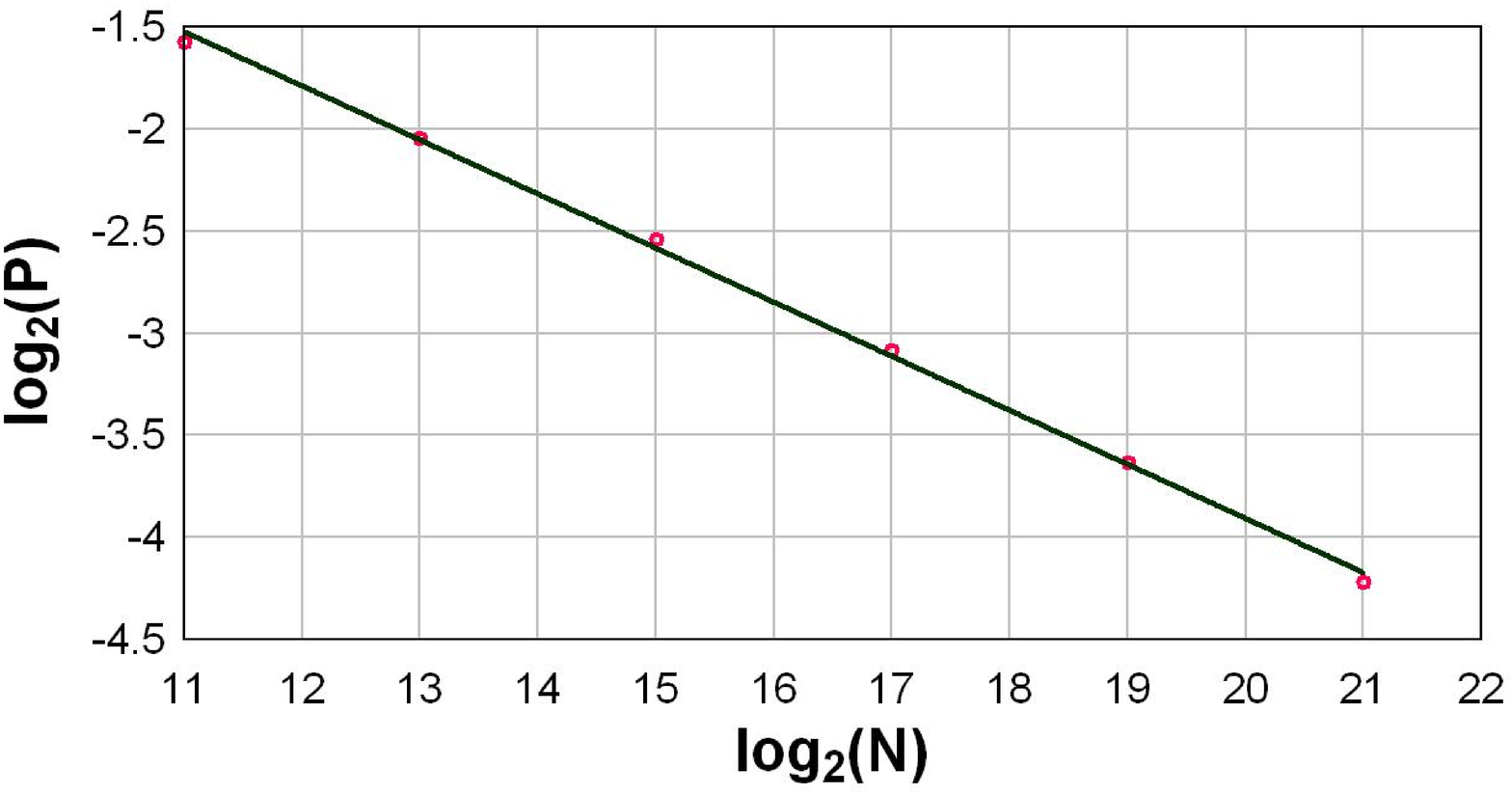}
\epsfxsize=8.8cm
\epsfbox{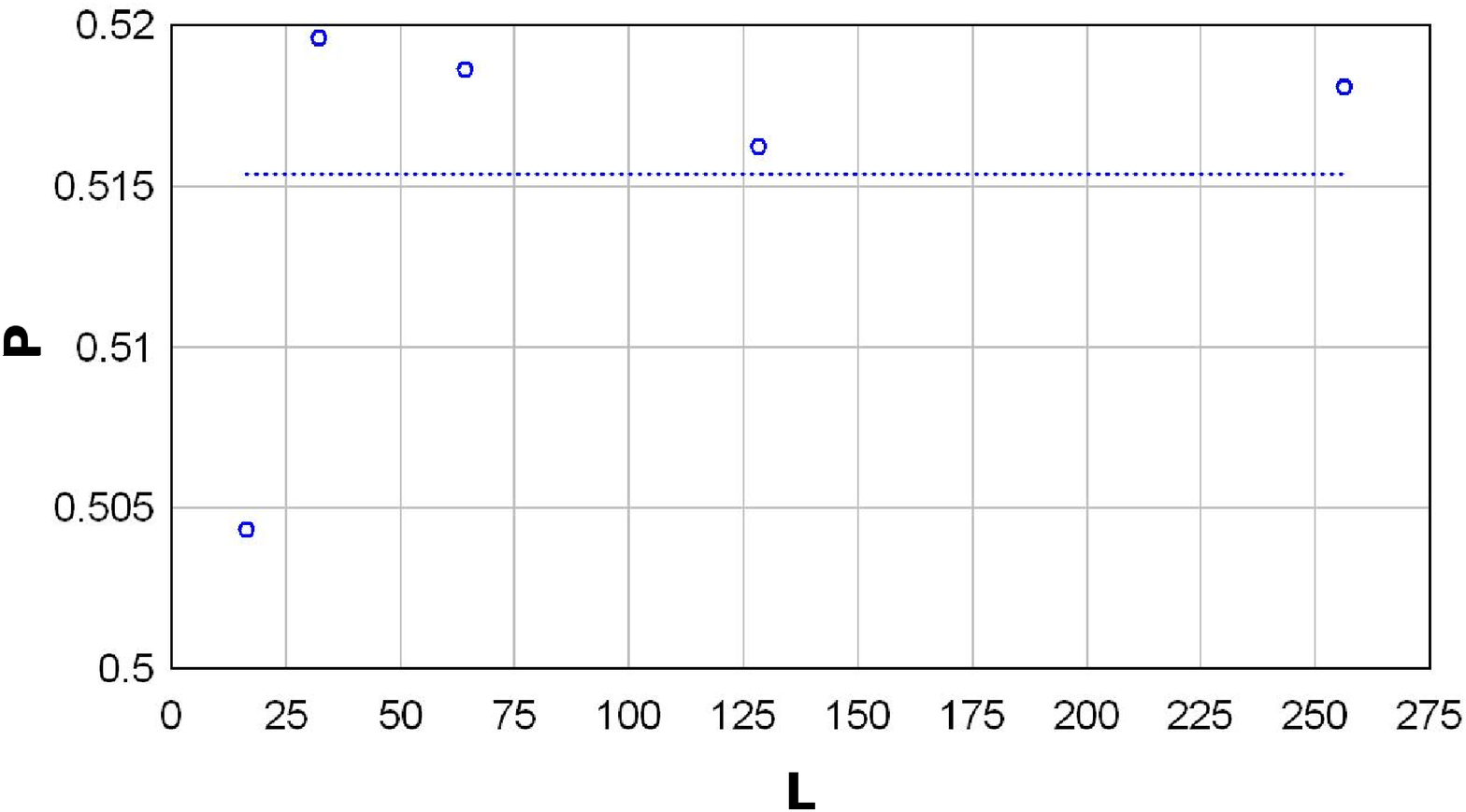}
\caption{(Color online) Scaling of the peak probability for the spatial search
on the three-dimensional Sierpinski gasket, without (top) and with (bottom)
ancilla control. The fits are for the data from $S=5$ to $S=10$, linear (top)
and constant (bottom) respectively.}
\end{figure}

With ancilla control, our fits yield:
\begin{eqnarray}
\log_2 Q_\delta &=& -0.817 + 0.647\log_2 N ~ (err=0.003), \cr
       P_\delta &=& 0.518 ~ (err=0.001).
\end{eqnarray}
We observe that $Q_\delta$ is only slightly larger than $Q_0$, the scaling
exponent $b$ is essentially the same, and $P_\delta$ is a constant close
to $0.5$, all consistent with Tulsi's predictions. We estimate the scaling
behavior as:
\begin{equation}
Q_\delta = 0.568(4) ~N^{0.647(1)} ~,~~ P_\delta = 0.518(1) ~.
\end{equation}
Moreover, direct fits for the complexity behavior give:
\begin{eqnarray}
\log_2 (Q_\delta/\sqrt{P_\delta}) &=& -0.343 + 0.647\log_2 N ~ (err=0.002), \cr
Q_\delta/\sqrt{P_\delta} &=& 0.788(3) ~N^{0.647(1)} ~.
\end{eqnarray}

Here the fact that $b\approx\frac{1}{d_s}$ is an important finding. In this
case, the embedding dimension $d_E=3$ and the Hausdorff dimension $d=2$.
Were either of them controlling the scaling behavior of the spatial search,
we would have found $Q=O(\sqrt{N})$. Our clearly distinct result once again
tells us that the dynamics of the spatial search is governed by the spectral
dimension.

\section{Conclusion}

We have described how the flip-flop walk contains within it a relativistic
Klein-Gordon propagator, and then used it to solve the spatial search problem
on fractals. The analogy with critical phenomena in statistical mechanics
suggests that, for $1<d<2$, physical observables have power-law scaling
behavior. Also, for $d\leq 2$, the fully relativistic spatial search algorithm
suffers from infrared divergence, which we have suppressed by Tulsi's ancilla
controlled version that is equivalent to the introduction of an effective mass
or a self-loop at the marked vertex.

Specifically, we have carried out numerical simulations for Sierpinski
gaskets in two and three dimensions, keeping the relative position of
the marked vertex in the gasket fixed, and our results support all the
expectations. An important finding is that we are able to reach the lower
bound of $N^{1/d}$ oracle calls, where $d$ is the spectral dimension and
not the fractal dimension. These two dimensions are not equal for fractals,
and we find that the smaller spectral dimension governs the scaling exponents.
Note that the solution of the quantum spatial search problem has to be
simultaneous in both the position space and the reciprocal momentum space.
The position space evolution proceeds from a uniform distribution toward
a $\delta$-function, and the momentum space evolution proceeds from a
$\delta$-function toward a uniform distribution. Thus it is fully
understandable that the stricter of the lower bounds (i.e. the smaller
value of $d_s$) governs the scaling behavior of the problem.


\begin{thebibliography}{99}
\bibitem{vielbein}{Note that the Clifford algebra cannot be formulated
in curved spaces either. Inclusion of fermions in general relativity
needs vielbeins, with one index in the curved space and the other in
the flat tangent space. The Dirac operator structure is defined in the
flat tangent space, and is contracted with curved space variables using
the vielbeins.}
\bibitem{firstorder}{A second-order differential equation can always be
rewritten as two coupled first-order differential equations, e.g.
$\ddot{\psi}+\psi=0$ and ${\displaystyle\frac{d}{dt}}
      \begin{pmatrix} \psi \cr \dot{\psi} \end{pmatrix}
      = \begin{pmatrix} 0 & 1 \cr -1 & 0 \end{pmatrix}
      \begin{pmatrix} \psi \cr \dot{\psi} \end{pmatrix}$
are equivalent descriptions of a scalar wave equation.}
\end{thebibliography}
\end{document}